\documentclass[prl,letterpaper,twocolumn,nofootinbib,showpacs]{revtex4}
\usepackage{amsfonts}
\usepackage{epsfig}

\newcommand{\Log}      	{{\mathrm {Log} \,}}

\newcommand{\be} 	{\begin{equation}}
\newcommand{\ee} 	{\end{equation}}
\newcommand{\ba} 	{\begin{eqnarray}}
\newcommand{\ea} 	{\end{eqnarray}}         
\newcommand{\nn} 	{\nonumber}

\newcommand{\Tr}	{\mathrm{Tr}}
\newcommand{\cI}	{{\cal I}}
\newcommand{\J}		{{\cal J}}
\newcommand{\Id}	{{\mathbf 1}}

\newcommand{\bra}[1]	{\left\langle #1 \right|}
\newcommand{\ket}[1]	{\left| #1 \right\rangle}
\newcommand{\cH}	{{\cal H}}
\newcommand{\cS}	{{\cal S}}
\newcommand{\cX}	{{\cal X}}
\newcommand{\cY}	{{\cal Y}}
\newcommand{\cA}	{{\cal A}}
\newcommand{\paragraf}[1] {{\it #1}}

\begin{document}

\title {Quantum discord: A measure of the quantumness of correlations}
    
\author{Harold Ollivier}
%\email[e-mail: ]{harold.ollivier@polytechnique.org}

\author{Wojciech H. Zurek}
%\email[e-mail: ]{whz@lanl.gov}
\affiliation{Theoretical Division T-6, MS B288, LANL, Los Alamos, NM 87545.} 


\begin{abstract}
Two classically identical expressions for the mutual information generally 
differ when the systems involved are quantum. This difference defines the
{\em quantum discord}. It can be used as a measure of the quantumness of 
correlations. Separability of the density matrix describing a pair 
of systems does not guarantee vanishing of the discord, thus showing that 
absence of entanglement does not imply classicality. We relate this 
to the quantum superposition principle, and consider the vanishing of discord 
as a criterion for the preferred effectively classical states of a system, 
i.e. the {\em pointer states}. 

\end{abstract}

\maketitle

%The original motivation for the pointer observable --- the observable whose
%eigenspaces, {\it pointer states}, are recorded by but do not entangle with
%the environment 
The original motivation for the pointer states  --- states that are monitored 
by the environment but do not entangle with it, and are therefore stable 
in spite of the openness of the system --- comes from the study of quantum
measurements \cite{bib:von neumann, bib:wheeler zurek, bib:peres}. When the 
quantum apparatus $\cA$ interacts with the system $\cS$ ({\em pre-measurement}),
the $\cS$--$\cA$ pair becomes entangled. The nature of the resulting quantum 
correlations makes it impossible to ascribe any independent {\em reality} to,
say, the state of the apparatus \cite{bib:zurek collapse}. For, a measurement 
of different observables on the state of the system will force the apparatus 
into mutually incompatible pure quantum states.
%In particular, its state cannot be ascertained without disturbing the state of the pair. 
This is a consequence of the {\em basis ambiguity}.  It is best exhibited by 
noting that the $\cS$--$\cA$ state after the pre-measurement,
\be \ket{\psi_{\cS,\cA}}^P = \sum_i\alpha_i\ket{s_i}\ket{a_i} \nn \ee
is typically entangled. One can rewrite it in a different basis of, 
e.g. the system, and one-to-one correlation with a corresponding set of pure, 
but not necessarily orthogonal, states of the apparatus will remain. 
Thus, it is obviously impossible to maintain that before the measurements 
the apparatus had a an unknown but real (i.e., existing independently of the 
system) quantum state.

Decoherence leads to environment-induced superselection (einselection) which 
singles out the pointer states and thus removes quantum excess of correlation 
responsible for the basis ambiguity. The density matrix of the decohering 
quantum apparatus loses its off-diagonal terms as a result of the interaction 
with the environment \cite{bib:zurek predsieve, bib:guilini et al, bib:paz zurek, bib:zurek rpm}:
\ba 
\rho_{\cS,\cA}^P & = & \ket{\psi_{\cS,\cA}}^P \bra{\psi_{\cS,\cA}}^P \nn \\* 
& \rightarrow & \sum_i |\alpha_i|^2 \ket{s_i}\bra{s_i}\otimes\ket{a_i}\bra{a_i} = \rho_{\cS,\cA}^D \label{einselection}.
\ea
Above $\bra{a_i}a_j\rangle = \delta_{i,j}$ following the ideal einselection process, which  transforms a pure $\rho_{\cS,\cA}^P$ into a decohered $\rho_{\cS,\cA}^D$ satisfying the superselection identity \cite{bib:bogolubov, bib:guilini}:
\be \rho_{\cS,\cA}^D = \sum_i P_i^\cA \rho_{\cS,\cA}^DP_i^\cA  \label{superselection}.\ee
Above $P_i^{\cA}$ correspond to the superselection sectors of the apparatus, 
e.g. the record states of its pointer (in our example 
$P_i^{\cA} = \ket{a_i}\bra{a_i}$).  One implication of this equation is that --- once einselection has forced the apparatus to abide by Eq.~(\ref{superselection}) ---
 its state can be consulted (measured) in the basis corresponding to the superselection sectors $P_i^{\cA}$ leaving $\rho_{\cS,\cA}^D$ unchanged \cite{bib:zurek predsieve, bib:zurek rpm}.

Einselection, Eq. (\ref{einselection}), obviously decreases correlations between $\cS$ and $\cA$. Yet, in a good measurement, one-to-one correlations between the pointer states of the apparatus and a corresponding set of system states must survive. We shall use two classically equivalent formulae for the mutual information to quantify the quantum and the classical strength of the correlations present in a joint density matrix $\rho_{\cS,\cA}$, and study the difference between these two as a measure of the quantum excess of correlations --- the quantum discord --- in $\rho_{\cS,\cA}$.

\paragraf{Mutual Information --- }
In classical information theory \cite{bib:cover} the entropy, $H(\cX)$, describes the ignorance about a random variable $\cX$, $H(\cX) = - \sum_x p_{\cX = x} \Log p_{\cX = x}$. The correlation between two random variables $\cX$ and $\cY$ is measured by the mutual information: 
\be \J(\cX:\cY) = H(\cX) - H(\cX|\cY) \label{Jclass}, \ee
where $H(\cX|\cY) = \sum_y p_{\cY=y} H(\cX|\cY = y)$ is the conditional entropy of $\cX$ given $\cY$. All the probability distributions are derived from the joint one, $p_{\cX,\cY}$: 
\ba 
&& p_\cX = \sum_y p_{\cX,\cY = y},\;  p_\cY = \sum_x p_{\cX=x,\cY} \\
&& p_{\cX|\cY = y} = {p_{\cX,\cY = y}}/{p_{\cY = y}} \quad\textrm{(Bayes rule)} \label{bayes}
\ea
Hence, the mutual information measures the average decrease of entropy on $\cX$ when $\cY$ is found out.
Using the Bayes rule, Eq.~(\ref{bayes}), one can show that $H(\cX|\cY) = H(\cX,\cY) - H(\cY)$. This leads to another classically equivalent expression 
for the mutual information:
\be \cI(\cX:\cY) = H(\cX) + H(\cY) -H(\cX,\cY) \label{Iclass}.\ee

One would like to generalize the concept of mutual information to quantum systems. 
One route to this goal, motivated by discussions of quantum information processing, has been put forward \cite{bib:schumacher nielsen, bib:cerf adami}. We shall pursue a different strategy, using Eqs.~(\ref{Jclass}) and (\ref{Iclass}). 
We start by defining $\cI$ and $\J$ for a pair of quantum systems.

\paragraf{\protect{$\cI$} --- }
All the ingredients involved in the definition of $\cI$ are easily generalized to deal with arbitrary quantum systems by replacing the classical probability distributions by the appropriate density matrices $\rho_\cS$, $\rho_\cA$ and $\rho_{\cS,\cA}$ and the Shannon entropy by the von Neumann entropy, e.g. $H(\cS) = H(\rho_\cS) = -\Tr_\cS \, \rho_\cS \Log \rho_\cS$:
\be \cI(S:A) = H(\cS) + H(\cA) - H(\cS,\cA).\ee
In this formula, $H(\cS) + H(\cA)$ represents the uncertainty of $\cS$ and $\cA$ treated separately, and $H(\cS,\cA)$ is the uncertainty about the combined system described by $\rho_{\cS,\cA}$. However, in contrast with the classical case, extracting all information potentially present in a combined quantum system described by $\rho_{\cS,\cA}$ will, in general, require a measurement on the combined Hilbert space $\cH_{\cS,\cA} = \cH_\cS \otimes \cH_\cA$.
The quantum version of $\cI$ has been used some years ago to study entanglement \cite{bib:zurek optics}, and subsequently rediscovered \cite{bib:Barnett Phoenix}.

\paragraf{\protect{$\J$} --- }
The generalization of this expression is not as automatic as for $\cI$, since the conditional entropy $H(\cS|\cA)$ requires us to specify the state of $\cS$ {\it given} the state of $\cA$. Such statement in quantum theory is ambiguous until the to-be-measured set of states $\cA$ is selected. 
We focus on perfect measurements of $\cA$ defined by a set of one-dimensional projectors $\{\Pi_j^\cA\}$. The label $j$ distinguishes different outcomes of this measurement.

The state of $\cS$, after the outcome corresponding to $\Pi_j^\cA$ has been detected, is 
\be \rho_{\cS|\Pi_j^\cA} = {\Pi_j^\cA\rho_{\cS,\cA}\Pi_j^\cA}/{\Tr_{\cS,\cA} \, \Pi_j^\cA\rho_{\cS,\cA}}, \ee
with probability $p_j = \Tr_{\cS,\cA} \, \Pi_j^\cA\rho_{\cS,\cA}$. 
$H(\rho_{\cS|\Pi_j^\cA})$ is the missing information about $\cS$. 
The entropies $H(\rho_{\cS|\Pi_j^\cA})$, weighted by probabilities, $p_j$, yield to the conditional entropy of $\cS$ given the complete measurement $\{\Pi_j^\cA\}$ on $\cA$,
\be H(\cS|\{\Pi_j^\cA\}) = \sum_j p_j \, H(\rho_{\cS|\Pi_j^\cA}).\ee
This leads to the following quantum generalization of $\J$:
\be \J(\cS:\cA)_{\{\Pi_j^\cA\}} = H(\cS) - H(\cS|\{\Pi_j^\cA\}).\ee
This quantity represents the information gained about the system $\cS$ as a result of the measurement $\{\Pi_j^\cA\}$.

\paragraf{Quantum discord --- } 
The two classically identical expressions for the mutual information, Eqs.~(\ref{Jclass}) and (\ref{Iclass}), differ in a quantum case \cite{bib:zurek discord}. The quantum discord is this difference,
\ba 
\delta(\cS:\cA)_{\{\Pi_j^\cA\}} & = & \cI(\cS:\cA) - \J(\cS:\cA)_{\{\Pi_j^\cA\}} \\
& = & H(\cA) - H(\cS,\cA) + H(\cS|\{\Pi_j^\cA\}).
\ea
It depends both on $\rho_{\cS,\cA}$ and on the projectors $\{\Pi_j^\cA\}$. 

The quantum discord is asymmetric under the change $\cS \leftrightarrow \cA$ since the definition of the conditional entropy $H(\cS|\{\Pi_j^\cA\})$ involves a measurement on one end (in our case the apparatus $\cA$), that allows the observer to infer the state of $\cS$. This typically involves an increase of entropy. Hence $H(\cS|\{\Pi_j^\cA\}) \geq H(\cS,\cA) - H(\cA)$, which implies that for any measurement $\{\Pi_j^\cA\}$,
\be \delta(\cS:\cA)_{\{\Pi_j^\cA\}} \geq 0.\ee
The proofs are postponed to the end of this letter.

We shall be usually concerned about the set $\{\Pi_j^\cA\}$ that minimizes the discord given a certain $\rho_{\cS,\cA}$. Minimizing the discord over the possible measurements on $\cA$ corresponds to finding the measurement that disturbs least the overall quantum state and that, at the same time, allows one to 
extract the most information about $\cS$. Decoherence picks out a set of stable states and converts their possible superpositions into mixtures, Eq.~(\ref{einselection}). Moreover, an unread measurement  
$\{\Pi_j^\cA\}$ on the apparatus has an effect analogous to einselection in 
the corresponding basis through the reduction postulate \cite{bib:von neumann}. 
Hence it is rather natural to expect that when the set $\{\Pi_j^\cA\}$ corresponds to the superselection sectors $\{P_i^\cA\}$ of Eq.~(\ref{superselection}), there would be no extra increase of entropy:
\be \rho_{\cS,\cA} = \sum_j \Pi_j^\cA \rho_{\cS,\cA} \Pi_j^\cA \Rightarrow \delta(\cS:\cA)_{\{\Pi_j^\cA\}} = 0 \label{ifthen}.\ee
Thus, following einselection, the information can be extracted from $\cS$--$\cA$ with a local measurement on $\cA$ without disturbing the overall state. The state of $\cS$ can be inferred from the outcome of the measurement on $\cA$ only. 
The converse of Eq.~(\ref{ifthen}) is also true:
\be \delta(\cS:\cA)_{\{\Pi_j^\cA\}} = 0 \Rightarrow \rho_{\cS,\cA} = \sum_j \Pi_j^\cA \rho_{\cS,\cA} \Pi_j^\cA.\ee
Hence, a vanishing discord can be considered as an indicator of the superselection rule, or --- in the case of interest --- its value is a measure of the efficiency of einselection. When $\delta$ is large for any set of projectors $\{\Pi_j^\cA\}$, a lot of information is missed and destroyed by any measurement on the apparatus alone, but when $\delta$ is small almost all the information about $\cS$ that exists in the $\cS$--$\cA$ correlations is locally recoverable from the state of the apparatus. 

\begin{figure}[htbp]
\epsfxsize 3.2in
\epsffile{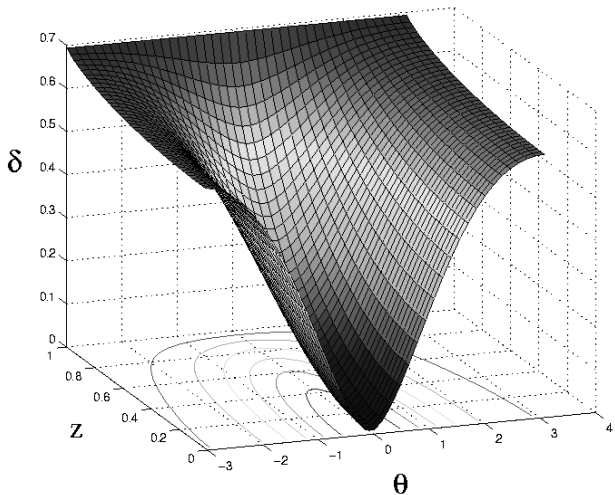}
\caption{Discord for the states given in \protect{Eq.~(\ref{state_p})}, with the measurement basis defined as in \protect{Eq.~(\ref{Pi})}.}% $\Pi_j(\theta) = U(\theta)\Pi_jU(\theta)^\dagger,\: \Pi_0 = \ket{0}\bra{0},\: \Pi_1 = \ket{1}\bra{1},\:  U(\theta) = \exp(i\theta(\cos(\varphi)\sigma_y + \sin(\varphi)\sigma_z)), \varphi = 1\textrm{rad}$.}
\label{fig:cnot}
\end{figure}

The quantum discord can be illustrated in a simple model of measurement. Let us
assume the initial state of $\cS$ is $(\ket{0} + \ket{1})/\sqrt{2}$. The pre-measurement is a {\tt c-not} gate yielding $\ket{\psi_{\cS,\cA}}^P = (\ket{00} + \ket{11})/\sqrt{2}$. If $\ket{0}$ and $\ket{1}$ of $\cA$ are pointer states, 
partial decoherence will suppress off-diagonal terms of the density matrix:
\ba \rho_{\cS,\cA} & = & \frac{1}{2}\left(\ket{00}\bra{00} + \ket{11}\bra{11} \right)  \nn \\
& & +\,  \frac{z}{2}\left(\ket{00}\bra{11} + \ket{11}\bra{00}\right) \label{state_p},
\ea
with $0\leq z <1$. Fig.~\ref{fig:cnot} shows $\delta$ for various values of $z$ and various bases of measurement parametrized by $\theta$,
\be 
\{\cos(\theta)\ket{0} + e^{i\phi}\sin{\theta}\ket{1}, 
e^{-i\phi}\sin{\theta}\ket{0} -\cos{\theta} \ket{1} \label{Pi}\},
\ee
with $\phi = 1{\textrm{rad}}$. %More precisely the measurement is composed by the one-dimensional projectors $\Pi_j(\theta) = U(\theta)\Pi_jU(\theta)^\dagger,\: \Pi_0 = \ket{0}\bra{0},\: \Pi_1 = \ket{1}\bra{1},\:  U(\theta) = \exp(i\theta(\cos(\varphi)\sigma_y + \sin(\varphi)\sigma_z)), \varphi = 1\textrm{rad}$.
Only in the case of complete einselection ($z=0$) there exist a basis for
which discord disappears. The corresponding basis of measurement is $\{\ket{0}, \ket{1}\}$ ($\theta = 0$), i.e. it must be carried out in the pointer basis.

\paragraf{Classical aspect of quantum correlations ---}
Separability has been often regarded as synonymous of classicality. The temptation that leads one to this conclusion starts with an observation that --- by definition --- a separable density matrix is a mixture of density matrices 
\be \rho_{\cS,\cA} = \sum_i p_i \rho^i_{\cS,\cA} \label{separable1} \ee
that have explicit product eigenstates, 
\be  \rho^i_{\cS,\cA} = \sum_j p_j^{(i)} \ket{s_j^{(i)}}\ket{a_j^{(i)}}\bra{a_j^{(i)}}\bra{s_j^{(i)}} \label{separable2}\ee 
and hence classical correlations. One might have thought that mixing such obviously classical density matrices cannot bring in anything quantum: After all, it involves only loss of information --- forgetting of the label $i$ in $\rho_{\cS,\cA}^i$. Yet this is not the case. One symptom of the quantumness of a separable $\rho_{\cS,\cA}$ with non-zero discord is immediately apparent: Unless there exists a complete set of projectors $\{\Pi_j^\cA\}$ for which $\delta(\cS:\cA)_{\{\Pi_j^\cA\}} = 0$, $\rho_{\cS,\cA}$ is perturbed by {\em all} local measurements. By contrast, when $\delta(\cS:\cA)_{\{\Pi_j^\cA\}} = 0$, then the measurement $\{\Pi_j^\cA\}$ on $\cA$, and an appropriate conditional measurement (i.e. conditioned by the outcome of the measurement on $\cA$) will reveal all of the information in $\cS$--$\cA$, i.e. the resulting state of the pair will be pure. Moreover this procedure can be accomplished without perturbing the $\rho_{\cS,\cA}$ for another observer, a bystande!
r not aware of the outcomes. 

Thus, for each outcome $j$ there  exist a set $\{\pi_{j,k}^\cS\}$ of conditional one dimensional projectors such that 
\be \rho_{\cS,\cA} = \sum_j \sum_k \pi_{j,k}^\cS\Pi_j^\cA\rho_{\cS,\cA}\Pi_j^\cA\pi_{j,k}^\cS, \ee
and $\pi_{j,k}^\cS\Pi_j^\cA\rho_{\cS,\cA}\Pi_j^\cA\pi_{j,k}^\cS$ is pure for any $j$ and $k$. Above, the sets  $\{\pi_{j,k}^\cS\}$ for different $j$ will not coincide in general ($\{\pi_{j,k}^\cS\}$ is a function of $j$) and do not need to commute. 
{\it Classical information is locally accessible, and can be obtained without perturbing the state of the system}: One can interrogate just one part of 
a composite system and discover its state while leaving the overall density 
matrix (as perceived by observers that do not have access to the measurement 
outcome) unaltered. 
A general separable $\rho_{\cS,\cA}$ does not allow for such insensitivity 
to measurements: Information can be extracted from the apparatus but only at 
a price of perturbing $\rho_{\cS, \cA}$, even when this density matrix is
separable.  However, when discord disappears, such insensitivity 
(which may be the defining feature of ``classical reality'', as it allowes 
acquisition of information without perturbation of the underlying state)
becomes possible for correlated quantum systems.
This quantum character of separable density matrices with non zero discord is a consequence of the superposition principle for $\cA$, since more than one basis $\left\{\ket{a_j^{(i)}}\right\}_j$ for the apparatus is needed in Eq.~(\ref{separable2}) in order to warrant a non vanishing discord.

\begin{figure}[htbp]
\epsfxsize 3.2in
\epsffile{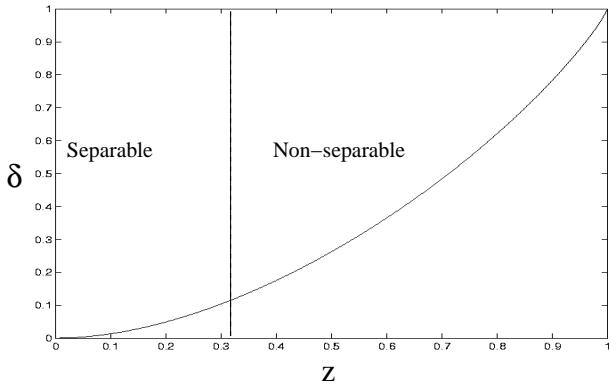}
\caption{Value of the discord for Werner states \protect{$\frac{1-z}{4}\Id + z\ket{\psi}\bra{\psi}$}, with \protect{$\ket{\psi} =(\ket{00}+\ket{11})/\sqrt{2}$}. Discord does not depend on the basis of measurement in this case because both \protect{$\Id$} and \protect{$\ket{\psi}$} are invariant under rotations.}
\label{fig:werner}
\end{figure}

The difference between separability and vanishing discord can be illustrated by a specific example. Fig.~\ref{fig:werner} shows discord 
for a Werner state $\rho_{\cS,\cA} = \frac{1-z}{4}\Id + z\ket{\psi}\bra{\psi}$ with $\ket{\psi} =(\ket{00}+\ket{11})/\sqrt{2}$. It can be seen that discord is greater than 0 in any basis when $z>0$, which contrasts with the well-known separability of such states when $z < 1/3$.  

\paragraf{Conclusion --- }
The quantum discord is a measure of the information that cannot be extracted by the reading of the state of the apparatus (i.e. without joint measurements). Hence the quantum discord is a good indicator of the quantum nature of the correlations. The pointer states obtained by minimizing the quantum discord over the possible measurements should coincide with the ones obtained with the predictability sieve criterion \cite{bib:zurek predsieve, bib:paz zurek}, hence showing that the accessible information remains in the most stable {\em pointer} states.

%It reinforces the statement obtained with the predictability sieve criterion \cite{bib:zurek predsieve, bib:paz zurek}: the accessible information is the one remaining in the most stable {\em pointer} states.

\begin{center} --- \end{center}

\paragraf{Proposition 1.}
$H(\cS|\{\Pi_j^\cA\}) = H(\rho_{\cS,\cA}^D) - H(\rho_\cA^D)$, with $\rho_{\cS,\cA}^D = \sum_j \Pi_j^\cA \rho_{\cS,\cA} \Pi_j^\cA$.

\paragraf{Proof 1.}
$\rho_{\cS,\cA}^D$ is block-diagonal. The $j$-th block equals $p_j \rho_{\cS|\Pi_j^\cA}$. By doing calculations block by block one has:

$H(\rho_{\cS,\cA}^D)  = \sum_j H( p_j\rho_{\cS|\Pi_j^\cA}) 
 =  \sum_j p_j H(\rho_{\cS|\Pi_j^\cA}) - \sum_j p_j \Log p_j 
 =  H(\cS|\{\Pi_j^\cA\}) - H(\rho_\cA^D),$ which completes the proof.

\paragraf{Proposition 2.}
$\delta(\cS:\cA)_{\{\Pi_j^\cA\}} \geq 0$.

\paragraf{Proof 2.}
This is a direct consequence of the previous proposition and the concavity of $H(\cS,\cA) - H(\cA)$ \cite{bib:wehrl}.

\paragraf{Proposition 3.}
$\delta(\cS:\cA)_{\{\Pi_j^\cA\}} = 0 \Leftrightarrow \rho_{\cS,\cA} = \sum_j \Pi_j^\cA \rho_{\cS,\cA} \Pi_j^\cA$.

\paragraf{Proof 3.}
Proposition 1 already shows the converse. To prove the direct implication we will start with $\rho_{\cS,\cA}$ and $\{\Pi_j^\cA\}$, a complete set of orthogonal projectors, such that $\delta(\cS:\cA)_{\{\Pi_j^\cA\}} = 0$. 
Without loss of generality, we can write the density matrix of $\cS$--$\cA$ as
$\rho_{\cS,\cA} = \sum_j \Pi_j^\cA\rho_{\cS,\cA}\Pi_j^\cA + \mathrm{additional \; terms}.$
If we choose $\{\ket{s_i}\}$ a basis of $\cS$, and $\{\ket{a_{k|j}}\}_k$ a basis of the subspace of $\cA$ defined by $\Pi_j^\cA$, the general form of the additional terms in the above formula will be $c(s_i, s_{i'},a_{k|j}, a_{k'|j'})\ket{s_i}\bra{s_{i'}}\otimes\ket{a_{k|j}}\bra{a_{k'|j'}}$ with $j \neq j'$. Suppose that one of those coefficients is non-zero. By changing the basis $\{\ket{s_i}\}_i$, we can suppose $i \neq i'$. We introduce now a new density matrix $\hat \rho_{\cS,\cA}$ obtained from $\rho_{\cS,\cA}$ by removing the preceding matrix element and its complex conjugate. This ensures that $\hat \rho_{\cS,\cA}$ is associated with a physical state. This state satisfies
$ H(\hat\rho_{\cS,\cA}) > H(\rho_{\cS,\cA}) \mbox{ and }  H(\hat\rho_\cA) = H(\rho\cA).$
The concavity of $H(\cS,\cA) - H(\cA)$ implies inequalities; 
\ba 
H(\rho_{\cS,\cA}) - H(\rho_\cA) & < &  H(\hat\rho_{\cS,\cA}) -  H(\hat\rho_{\cA}), \nn \\
H(\hat\rho_{\cS,\cA}) - H(\hat\rho_{\cA}) & \leq & H(\rho_{\cS,\cA}^D) - H(\rho_{\cA}^D). \nn
\ea 

Then $\delta(\cS:\cA)_{\{\Pi_j^\cA\}} < 0$, which contradicts our primary assumption and proves our last result.

\paragraf{Remark.}
We defined $\J$ with the help of a measurement associated with one-dimensional projectors. One can be interested in looking at multi-dimensional projective measurements. Depending on the context, two different generalization can be used. 

For measurement purposes, one may adopt 
$$\rho_{\cS|\Pi_j^\cA} = {\Tr_\cA \, \Pi_j^\cA \rho_{\cS,\cA}}/{\Tr_{\cS,\cA} \, \Pi_j^\cA \rho_{\cS,\cA}}, $$
since all the correlations (quantum as well as classical) between $\cS$ and the subspace of the apparatus defined by $\Pi_j^\cA$ are not observed. Proposition 1 no longer holds, but using the same techniques we still have $\delta(\cS:\cA)_{\{\Pi_j^\cA\}} \geq 0$ and if $\delta(\cS:\cA)_{\{\Pi_j^\cA\}} = 0$, then $\rho_{\cS,\cA} = \sum_j \Pi_j^\cA \rho_{\cS,\cA} \Pi_j^\cA$.

For decoherence purposes, one may prefer to define $\J$ as $H(\cS) + H(\cA)^D - H(\cS,\cA)^D$. With this definition, Proposition 3 is valid. $\J$ now represents the average information, quantum and classical that remains in the pair $\cS$--$\cA$ after a decoherence process leading to einselection of the superselection sectors $\{\Pi_j^\cA\}$.

\begin{acknowledgments}
This research was supported in part by NSA. Preliminary results were presented at the Newton Institute in July 1999 by W.H.Z., at the Ushuaia PASI in October 2000 by H.O., and described in the proceedings of the $100^\mathrm{th}$ anniversary of Planck's Constant Meeting  \cite{bib:zurek discord}.  Useful conversations with E. Knill, R. Laflamme, B. Schumacher and L. Viola are gratefully acknowledged.
\end{acknowledgments}

\paragraf{Note ---} After completion of this work, we became aware of a related work by L\@. Henderson and V\@. Vedral ({\tt quant-ph/0105028}).

\end{document}